\RequirePackage{amsmath}
\documentclass[runningheads]{llncs}
\usepackage[T1]{fontenc}
\usepackage{graphicx}
\usepackage[misc,geometry]{ifsym}

\graphicspath{{figures/}}
\usepackage{color}

\usepackage{comment}
\usepackage{enumitem}
\usepackage{listings}
\usepackage{subcaption}
\usepackage[notransparent]{svg}
\usepackage{adjustbox}

\begin{document}
\title{Accelerating Drug Discovery in AutoDock-GPU with Tensor Cores}
\author{Gabin Schieffer \and Ivy Peng$^{\textrm{(\Letter)}}$}
\institute{KTH Royal Institute of Technology, Stockholm, Sweden \\ \email{\{gabins, ivybopeng\}@kth.se}}

\authorrunning{G. Schieffer and I. Peng}
\maketitle              %
\begin{abstract}
In drug discovery, molecular docking aims at characterizing the binding of a drug-like molecule to a macromolecule. AutoDock-GPU, a state-of-the-art docking software, estimates the geometrical conformation of a docked ligand-protein complex by minimizing a scoring function. Our profiling results indicate that the current reduction operation that is heavily used in the scoring function is sub-optimal. Thus, we developed a method to accelerate the sum reduction of four-element vectors using matrix operations on NVIDIA Tensor Cores. We integrated the new reduction operation into AutoDock-GPU and evaluated it on multiple chemical complexes on three GPUs. Our results show that our method for reduction operation is $4$-$7$ times faster than the AutoDock-GPU baseline. We also evaluated the impact of our method on the overall simulation time in the real-world docking simulation and achieved a 27\% improvement on the average docking time.

\end{abstract}

\section{Introduction}
The pharmacological effect of a drug is generally induced by the binding of a drug molecule to a specific protein target. Thus, characterizing the ability of binding is crucial for drug discovery. Once a target for a disease is identified, tens of millions of chemical compounds, or \textit{ligands}, will go through high-throughput screening. For such vast search space, virtual screening that leverages computational approaches is becoming increasingly important for accelerating the process and reducing the high cost required in experimental screenings~\cite{santos-martins_accelerating_2021,legrand_gpu-accelerated_2020}. In particular, structure-based virtual screening software uses molecular docking tools to test a molecule drug candidate for binding a protein target (receptor). In recent COVID-19 research, high-performance virtual screening software has been used in combating the pandemic~\cite{legrand_gpu-accelerated_2020}.

A typical molecular docking job consists of evaluating a large number of ligands, each as an independent docking task.
Further distributing individual docking tasks onto high-performance computing (HPC) systems, with multi-core CPU or GPUs, can significantly accelerate docking, e.g., AutoDock-GPU reports 350-fold speedup over single-threaded implementation~\cite{mermelstein2018fast,santos-martins_accelerating_2021}.
AutoDock is widely used in the pharmaceutical industry to characterize protein-ligand complexes. In recent efforts, AutoDock4 implements its search engine based on Lamarckian Genetic Algorithm (LGA) and is ported to GPUs. 
A CUDA implementation of AutoDock-GPU with enhanced workflow successfully scaled to leverage the Summit supercomputer~\cite{legrand_gpu-accelerated_2020}.

In this work, we focus on the CUDA implementation of AutoDock-GPU as it represents the state-of-the-art of docking software on HPC systems. AutoDock-GPU predicts the geometrical conformation of a ligand-protein complex by minimizing an energy-based \textit{scoring} function that quantifies the free energy of a given binding pose. A docking job typically have many LGA runs, each consisting of multiple iterations till reaching the max number of score evaluations or GA generations. Therefore, the scoring function is called many times, e.g., $10^6$ to $10^8$, in a docking job, dominating the runtime~\cite{santos-martins_accelerating_2021}. The scoring function parallelizes the computation of the energy and associated gradient values by distributing iterations across all threads in a block and computing the total energy in a block-level reduction operation. 
Our profiling results show that the current implementation of the reduction operation causes a significant proportion of the overall number of warp stalls in the local search kernel. 

We propose a Tensor Core based reduction operation to accelerate the docking process -- leveraging Tensor Core Units and reducing synchronization points. We designed a multi-dimensional reduction algorithm based on previous works~\cite{dakkak_accelerating_2019,navarro_gpu_2021}. Our design leverages compacted data layout in shared memory. By merging multiple matrix multiplications into a single one, we dramatically reduce the number of synchronization points. We implemented the new algorithm in CUDA using the Nvidia WMMA API and integrated it in the energy calculation function in AutoDock-GPU. We validated the implementation and then evaluated its performance in single kernel and overall docking time on three generations of NVIDIA GPUs, including T4, V100, and A100. The results show that our method consistently outperform the AutoDock-GPU baseline, achieving up to $6.7\times$ and $4.7\times$ speedup on A100 and V100, respectively. We summarize our contributions as follows:
\begin{itemize}[noitemsep,topsep=0pt]
    \item Our performance charaterization of the AutoDock-GPU identified the scalability bottleneck in reduction operation in scoring function
    \item We proposed a multi-dimension reduction operation leveraging the mixed-precision Tensor Core Units
    \item We provided an implementation in CUDA using WMMA API in AutoDock-GPU and validated the implementation
    \item We evaluated the performance within single kernel on three GPUs and achieved $4.1$-$6.7\times$ speedup, and a 27\% improvement on average docking time  
\end{itemize}

\section{Background}
\label{sec:bg}
In this section, we introduce the computation method in molecular docking and the GPU implementation of AutoDock-GPU. We also introduce Tensor Core Unit and its programming interfaces on NVIDIA GPUs. 

\subsection{Computational method in AutoDock-GPU}
\label{subsec:computational_method_autodock}
AutoDock~\cite{morris_automated_1998} variants, e.g., AutoDock-Vina, AutoDock4, and AutoDock-GPU, use an energy-based scoring function to measure the quality of a given binding pose. The scoring function is a free-energy force field. It captures contributions from various physical interactions between atom pairs to associate an energy value to a ligand-receptor conformation. Recent development~\cite{santos-martins_accelerating_2021} introduces different search algorithms, such as the Solis-Wets and the ADADELTA methods, to accelerate the docking.

In the docking method in AutoDock-GPU, the target molecule is fixed. Thus, the ligand-receptor complex can be fully described by a set of variables related to the position, rotation, and internal conformation of the ligand. This set of variables, referred as \textit{ligand pose} or \textit{genotype}, is composed of seven dimensions, i.e., $x,y,z$ representing the ligand's position in space, $\phi, \theta, \alpha$ characterizing the rotation of the ligand, and $N_{rot}$ dimensions characterizing the torsion angles of rotatable bonds in the ligand by $\psi_1 \ldots \psi_{N_{rot}}$. These variables are the input to the scoring function. %

AutoDock-GPU uses a parallelized version of the original LGA~\cite{santos-martins_accelerating_2021}. The LGA uses a genetic algorithm (GA) to perform a global search, which generates several genotypes (denoted as $\Omega$). Each genotype is then improved by a local search algorithm (LS) that minimizes the scoring function (free energy). Two commonly used local search algorithms are ADADELTA and Solis-Wets. ADADELTA~\cite{zeiler_adadelta_2012} is a gradient-based optimization algorithm. It updates the genotype $\Omega$ at each iteration $t$ by $\Omega_{t+1} = \Omega_t + \eta_t g_t$, where $\eta_t$ depends on the history of previous update and gradient values, and $g_t$ is the gradient of the scoring function at the point $\Omega_t$. The computational cost of this method is dominated by the gradient calculation.
AutoDock-GPU parallelizes computation of the energy value by distributing iterations across all threads in a block. Each thread computes a partial value of the total energy and a block-level reduction is used to compute the total energy value. Similarly, each thread computes a partial value of the gradient for each of the three geometrical dimensions $x, y, z$, as well as the torque generated by physical interactions on the ligand, which is required for the calculation of the rotation-related and torsion-related gradient values. In total, seven block-level reductions are required for each evaluation of the scoring function, during the local-search optimization process.

\subsection{{NVIDIA} Tensor Cores}
\label{sec:tcu}
NVIDIA Tensor Cores were introduced in the Volta GPU microarchitecture, providing tremendous computing power in reduced precision~\cite{markidis_nvidia_2018}. NVIDIA V100 features 640~first-generation Tensor Cores and a theoretical peak performance of 125~Tflops/s in mixed precision. The Turing architecture extended Tensor Cores abilities by adding support for computation using more data types. The Tesla T4 offers 320 Tensor Cores, and provides a theoretical peak performance of 65~Tflops/s. In the Ampere architecture, the A100 GPU features 432 Tensor Cores, and provides a theoretical peak performance of 312~Tflops/s.

Tensor Core Units (TCU) are designed to perform matrix multiply-and-accumulate operations (i.e., $V \leftarrow A \cdot B + V$) in high throughput, while enforcing constraints on matrix sizes and precision. The operands of the multiplication operation must be of size $16\times 16$ and contain half-precision elements~\cite{cuda_program}. The accumulator can use single-precision float representation.

Tensor Core operations use the \textit{half-precision} data type, which relies on a 16-bit binary representation. This level of precision is generally sufficient for deep learning workloads, and scientific workloads resilient to precision loss can also benefit from it. However, the \textit{half-precision} data type requires explicit conversion to the single-precision 32-bit float representation. Starting with the Ampere GPU architecture, NVIDIA added support for both \textit{bfloat16} and \textit{tf32} in Tensor Cores. While double-precision data type is also supported on Tensor Cores from the Ampere GPU architecture, the matrix size in this precision is limited to $8\times 4$ for the multiplication operands, and $8\times 8$ for the accumulator. %

The WMMA API (\textit{Warp Matrix Multiply-and-Add}) provides a limited set of functions for developers to use Tensor Cores. Codes using this API are portable across different NVIDIA GPU architecture. 
This API exposes functions to set up and perform multiply-and-accumulate operations on Tensor Cores.
It defines a data structure named \emph{fragment}. A fragment is an abstraction to represent a matrix. Each fragment holds the matrix metadata, i.e., the data type, the matrix size, and the type of matrix as either an operand or an accumulator. The actual matrix elements held by a fragment are spread across threads in the warp, this data-to-threads mapping is not known by the developer~\cite{dakkak_accelerating_2019}. Instead, the WMMA API provides basic load and store functions to map generic CUDA data structures, such as arrays, to fragments.
A multiply-and-accumulate operation is exposed as a function operating on fragments and requires the collaboration of all threads in a warp. %

\section{Performance Characterization on GPU}
\label{sec:profiling}
\begin{figure}[ht]
    \centering
    \includegraphics[width=0.9\textwidth]{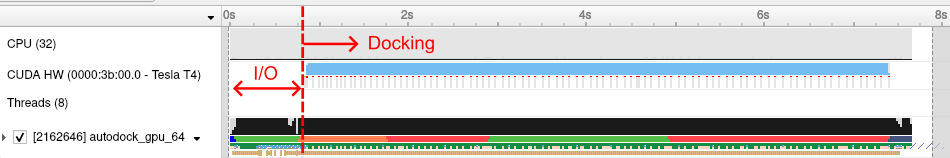}
    \caption{Profiling results of a docking process of the \texttt{7cpa} protein-ligand complex.}
    \label{fig:method_nsys_full_timeline}
\end{figure}

In this section, we first provide an overview of the runtime breakdown of a simulation and then focus on the GPU computation.
We used the \texttt{7cpa} protein-ligand complex and ran with a block size of 64 threads on NVIDIA A100 GPU, using all default parameters. The profiling results were obtained with NVIDIA Nsight Systems. At high level, the runtime of a simulation is dominated by the docking time, which is GPU bound, and then I/O pre-processing~\cite{markidis2021understanding}.
In Fig.~\ref{fig:method_nsys_full_timeline}, NVIDIA Nsight Systems reports $90\%$ time spent in docking.

In the docking process, the runtime is dominated by the local-search kernel, \lstinline{gpu_gradient_minAD}.
As shown in Fig.~\ref{fig:method_nsys_kernels_only}, the gradient-based local search dominates the docking time on GPU, i.e., $99.6\%$ kernel time is spent in the \lstinline{gpu_gradient_minAD} kernel (the details are described in ~\cite{santos-martins_accelerating_2021}). The breakdown of GPU kernel runtime is reported in Table~\ref{tab:method_kernel_percents}.
In this kernel, seven reduction operations are performed to compute the value and gradient of the scoring function, which happens at every iteration of the gradient-descent algorithm. This reduction operation is defined as a C++ macro named \lstinline{REDUCEFLOATSUM} (denoted as \textit{ReduceFS} in the remainder of this paper).

\begin{table}[bt]
\begin{minipage}[t]{0.48\linewidth}
\centering
    \includegraphics[width=\textwidth]{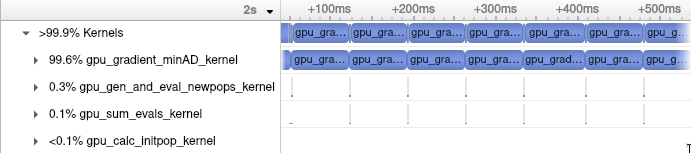}
    \captionof{figure}{The kernel launch timeline for iterations of the optimization process.}
    \label{fig:method_nsys_kernels_only}
\end{minipage}
~~
\begin{minipage}[t]{0.48\linewidth}
\centering
\resizebox{\textwidth}{!}{%
    \begin{tabular}[b]{l|c}
        kernel name  & \% of total kernel runtime \\
        \hline\hline
        \lstinline|gpu_calc_initprop_kernel|        & <0.1\% \\
        \lstinline|gpu_sum_evals_kernel|            & 0.1\% \\
        \lstinline|gpu_gen_and_eval_newpops_kernel| & 0.3\% \\
        \lstinline|gpu_gradient_minAD_kernel|       & 99.6\% \\
    \end{tabular}		
  }
\caption{Time breakdown in CUDA kernels}
\label{tab:method_kernel_percents}
\end{minipage}
\end{table}

\begin{figure}[bt]
    \centering
    \includegraphics[width=0.9\textwidth]{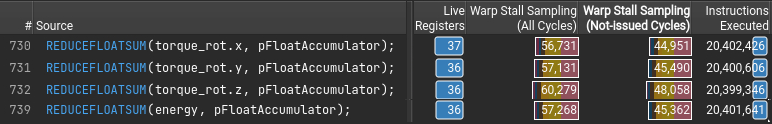}
    \caption{Profiling results of the \lstinline|gpu_gradient_minAD_kernel| kernel.}
    \label{fig:method_ncu_source}
\end{figure}

We observe a large number of warp stalls in each execution of \textit{ReduceFS} in Fig.~\ref{fig:method_ncu_source}, which reports four consecutive calls of \textit{ReduceFS} macro. Moreover, these lines of code are identified among the top ten lines of code causing high numbers of warp stalls, indicating that the stalls could have a high impact on overall kernel performance. From the causes for these warp stalls returned by NVIDIA Nsight Compute, we observe that approximately 40\% of warp stalls are caused by memory barriers (``membar''), related to the use of memory fence operations. Also, about 25\% of warp stalls are caused by ``short scoreboard'', which is often caused by shared memory instruction latency.

The profiling results led us to investigate further the block-level reduction in AutoDock-GPU. We established that \lstinline{REDUCEFLOATSUM(value, acc)} performs a block-level reduce-and-broadcast operation. Each thread provides one single-precision number \textit{value}, which will be reduced with all other values for other threads in the block. At the end of the reduction, the result is placed back in \textit{value}. \textit{acc} is a pointer to a \lstinline{float} in shared memory, which is used internally as an accumulator to perform reduction.

The current implementation mainly relies on three CUDA functions -- warp shuffle functions, atomic operations, and block-level synchronizations. First, a warp-level reduction is performed through warp shuffle functions, which allow data exchange between threads within a warp without using shared memory. In particular, the \lstinline{__shfl_sync} function allows a thread to read a value from another thread within the same warp, in a synchronized fashion.

In the warp-level reduction algorithm, this function is called multiple times by each thread. At each call, each thread adds the value received from another thread into its local copy. By organizing communication in a tree-like pattern, five consecutive calls to \lstinline{__shfl_sync} are sufficient for each thread to have its own local copy of the total sum across all 32 threads (a warp). This warp-level reduction algorithm is state-of-the-art~\cite{dakkak_accelerating_2019}.

After the warp-level reduction is completed, the first thread of each warp performs an atomic add of the result to a shared memory accumulator. Finally, each thread in a thread block performs a read from the accumulator in shared memory to receive the reduction result, finishing the whole operation. 

\noindent\textit{Takeaway 1: Atomic operations are used for value accumulation, and could cause contention when a large number of warps is used.}

As described in Section~\ref{sec:bg}, the scoring function implementation needs to perform reduction over seven dimensions -- one for the global energy value, three for the gradient calculation, and three for the torque calculation.
In the current AutoDock-GPU version, this is implemented by sequentially calling the \textit{ReduceFS} macro seven times in the scoring function kernel.

\noindent\textit{Takeaway 2: each evaluation of the scoring function repeats the block-level reduction operation seven times sequentially.}

For each use of \textit{ReduceFS}, three explicit block-level thread synchronizations are performed, which results in a total of 21 synchronizations for the seven-dimensional reduction. This could drastically reduce the parallelism of the algorithm. 

\noindent\textit{Takeaway 3: Performing reduction operation on seven dimensions separately results in 21 block-level synchronizations, a potential bottleneck for scalability.}

\section{Methodology}
In this work, we leverage Tensor Core Units (TCU) to accelerate matrix-based reduction. In~\cite{dakkak_accelerating_2019}, scan and reduction operations on an array are expressed as matrix operations and accelerated on NVIDIA Tensor Cores. This method relies on placing the elements to be reduced in a matrix, which is then multiplied by a well-chosen matrix to perform summation on the rows. A similar operation is then applied to perform summation on the columns. This line-then-column summation process effectively sum up all elements, equivalent to performing a reduction operation.

We propose an approach to replace the reduction operation in AutoDock-GPU by an implementation of a reduction method which is able to leverage Tensor Core Units. We first list the requirements that our method must meet to be used in AutoDock-GPU code. Then, we describe how we adapt and optimize the general Tensor Core-based reduction operation to meet the specific requirements in AutoDock-GPU. It is worth noting here that even though the method and implementation proposed in this paper are tailored to a specific application, the performed operation is general. Therefore, our approach can be generalized to other applications, with reasonable adaptation efforts.

\subsection{Requirements and design choices}
The scoring function in AutoDock-GPU performs seven consecutive reductions, each time for one variable. Previous TCU-based reduction method only reduces one variable at a time. To improve the efficiency, we propose to merge the reduction operations of four variables. This change would bring two main benefits. First, the profiling results show that a single reduction operation inherently requires synchronization between threads. Thus, merging four reductions would ideally reduce the synchronization cost by four times, improving parallelism. Second, we can improve the efficiency of data movement by reducing the number of separate data transfers. As introduced in Section~\ref{sec:tcu}, data arrays needs to be transferred (and mapped) from shared memory to be used on TCUs. By transforming the data layout into one contiguous data layout in shared memory, this overhead can be reduced.

The mapping between matrix elements and thread registers is not consistent across different GPU architectures. For this reason, NVIDIA recommends using the exposed API functions, i.e., \lstinline{load_matrix_sync()} to load matrices data. When this function is called, each thread copies a portion of shared memory array to its registers. The matrix data is hence spread across all threads in the warp. This process may be sub-optimal in applications where matrices elements are already initially stored in registers, since those elements would first need to be copied to shared memory and then loaded to registers while they only need to be read back from registers. For this reason, previous work~\cite{jia_dissecting_2018} has reverse-engineered the memory mapping between matrix elements and corresponding thread registers. Previous TCU-based reduction method~\cite{dakkak_accelerating_2019} chose to use this knowledge to manipulate matrix data directly in registers. 

In AutoDock-GPU, matrix elements are initially stored in each thread's registers. Thus, the reverse engineered memory mapping technique could squeeze more performance. However, this technique also requires specific tuning for each architecture. Therefore, for portability across different GPUs, we chose to use the NVIDIA-recommended approach. %

\subsection{Matrix-based multi-dimensional reduction method}
\label{subsec:matrix_ops}
We design a method using matrix operations to perform sum reduction of a set of four-element vectors. Our method aims at computing the sum of $n$ four-element vectors $\vec{u_i}=(x_i, y_i, z_i, e_i)$. The result is also a four-element vector, which contains on each of its coordinates the sum for each corresponding dimension, i.e., $\vec{y_i}=(\sum_i x_i, \sum_i y_i, \sum_i z_i, \sum_i e_i)$. We represent our input data as a $16\times 16$ matrix $A$, containing coordinates of the first $64$ vectors, organized in a column-major fashion. We also declare two $16\times 16$ matrices -- $P$ and $Q$. $P$ is a matrix filled with ones. $Q$ is a block-matrix composed of $4\times 4$ blocks, each being the $4\times 4$ identity matrix $I_4$.

\begin{minipage}{.35\linewidth}
$A = \begin{pmatrix}
x_0    & x_4    & \dots & x_{60} \\
y_0    & y_4    & \dots & y_{60} \\
z_0    & z_4    & \dots & z_{60} \\
e_0    & e_4    & \dots & e_{60} \\
\vdots & \vdots &       & \vdots
\end{pmatrix}$
\end{minipage}%
\begin{minipage}{.3\linewidth}
$P = \begin{pmatrix}
1      & \dots  & 1 \\
\vdots & \ddots & \vdots \\
1      & \dots  & 1
\end{pmatrix}$
\end{minipage}
\begin{minipage}{.3\linewidth}
$Q = \begin{pmatrix}
I_4 & I_4 & I_4 & I_4 \\
I_4 & I_4 & I_4 & I_4 \\
I_4 & I_4 & I_4 & I_4 \\
I_4 & I_4 & I_4 & I_4
\end{pmatrix}$
\end{minipage}

We first compute the matrix product $AP$ into $V$. This operation effectively performs summation on the rows. If more than 64 vectors need to be reduced, we iterate the same operation, each time with $A$ containing elements for a new set of 64 vectors in the input dataset and accumulating the results into $V$. We then perform sum on every 4\textsuperscript{th} column in $V$ with the matrix operation $QV$ and save the result into $W$. At this point, the matrix $W$ contains the desired result as the four first elements on the first column.

\begin{minipage}{.57\linewidth}
$V \leftarrow AP = \begin{pmatrix} %
\sum x_{4i} & \sum x_{4i} & \dots & \sum x_{4i} \\
\sum y_{4i} & \sum y_{4i} & \dots & \sum y_{4i} \\
\sum z_{4i} & \sum z_{4i} & \dots & \sum z_{4i} \\
\sum e_{4i} & \sum e_{4i} & \dots & \sum e_{4i} \\
\sum x_{4i+1} & \sum x_{4i+1} & \dots & \sum x_{4i+1} \\
\sum y_{4i+1} & \sum y_{4i+1} & \dots & \sum y_{4i+1} \\
\sum z_{4i+1} & \sum z_{4i+1} & \dots & \sum z_{4i+1} \\
\sum e_{4i+1} & \sum e_{4i+1} & \dots & \sum e_{4i+1} \\
\vdots & \vdots & \vdots & \vdots \\
\end{pmatrix}$
\end{minipage}~~~~~~
\begin{minipage}{.42\linewidth}
$V \leftarrow AP + V$\\
\\
$W \leftarrow QV$ \\
\\
$W = \begin{pmatrix}
\sum x_{i} & \sum x_{i} & \dots & \sum x_{i} \\
\sum y_{i} & \sum y_{i} & \dots & \sum y_{i} \\
\sum z_{i} & \sum z_{i} & \dots & \sum z_{i} \\
\sum e_{i} & \sum e_{i} & \dots & \sum e_{i} \\
\vdots & \vdots & \vdots & \vdots \\
\end{pmatrix}$
\end{minipage}

We implement our method as a CUDA \lstinline{__device__} function using the NVIDIA WMMA API to perform matrix operations. This function replaces four sequential uses of the \textit{ReduceFS} macro in the energy-and-gradient calculation in AutoDock-GPU. The four elements to be reduced for each thread are first converted from float to half-precision using the CUDA \lstinline{half2float} function, and then loaded into a contiguous data array in shared memory. The data loading is collectively performed by all threads in a block.

The accumulator $V$ is a product of matrices $A$ and $P$. Meanwhile, it is also an operand for the matrix multiplication calculating $W$. Then, in order to compute $W$ using TCUs, $V$ must be half-precision.
Using single precision for accumulation in $V$ would require to convert it to half-precision before computing $W$, a casting back to single precision would then be necessary. This approach requires two non-trivial conversions between two levels of precision. Instead, we choose to use half-precision for both operations. 

In our implementation, 
two block-level synchronizations are needed in total. A first one is performed before the first WMMA API call, to ensure that values for all threads are available in shared memory before starting the reduction process. The second synchronization is performed after the last WMMA API call, to ensure that all threads in the block can read the results. Compared to the $21$ synchronizations in original AutoDock-GPU, our method significantly reduces synchronization points.

Our implementation requires no memory barriers and atomic operations, unlike the current AutoDock-GPU method. Note that those operations are responsible for a significant number of stalls (Section~\ref{sec:profiling}). In addition, the decreased amount of those contention-causing operations could improve scalability.

\section{Evaluation}
We evaluated our implementation on four testbeds, featuring three GPU architectures, i.e., T4, V100, and A100. We summarize their system specifications in Table~\ref{tab:testbed}. %
Docking experiments were performed using five protein-ligand complexes, referred by their four-character Protein Data Bank identifier. We used the following complexes: \texttt{1stp}, \texttt{7cpa}, \texttt{1ac8}, \texttt{3tmn}, \texttt{3ce3}. Those five complexes, which are real-world samples, are provided with AutoDock-GPU code as test samples. Three of them were chosen for their particular molecular characteristics, in order to validate various aspects of the docking implementation, in particular the gradient calculation.

\begin{table}[b]
\caption{A summary of four testbeds used for evaluation}
\label{tab:testbed}
\begin{adjustbox}{width=\textwidth,center}{
\centering
\begin{tabular}{|c|c|c|c|c|c|c}
 \hline
 \textbf{Testbed} &\textbf{GPU} &\textbf{CPU} &\textbf{Interconnect} &\textbf{GPU Memory} &\textbf{CPU Memory}\\\hline\hline
TB1 &NVIDIA Tesla T4  &16 core Intel(R) Xeon(R) Gold &PCIe &16GB RAM &576GB DDR4\\
TB2 &NVIDIA Tesla V100 SXM2  &8 core Intel(R) Xeon(R) Gold &NVLink &32GB HBM2 &768GB DDR4\\
TB3 &NVIDIA Tesla V100 SXM2  &16 core Intel(R) Xeon(R) Gold &NVLink &32GB HBM2 &768GB DDR4\\
TB4 &NVIDIA Tesla A100  &32 core Intel(R) Xeon(R) Gold &NVLink &40GB HBM2 &576GB DDR4\\\hline\hline
\end{tabular}
}\end{adjustbox}
\end{table}

\subsection{Validation of the Scoring Function}
\label{sec:validation}
Our first step is to validate the TCU-based implementation in AutoDock-GPU scoring function. For this, we leverage similar metrics defined in ~\cite{santos-martins_accelerating_2021} to evaluate the correctness in LGA run and overall simulations. In particular, we compare simulation results to the baseline results to quantify the precision loss introduced by the half-precision operations on TCU.

\begin{figure}[bt]
    \centering
    \begin{tabular*}{\linewidth}{@{\extracolsep{\fill}} ccccc } 
    \multicolumn{5}{c}{\includegraphics[height=4mm]{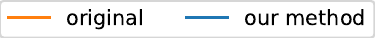}} \\
    \includegraphics[height=33mm]{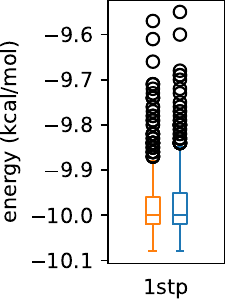} & %
    \includegraphics[height=33mm]{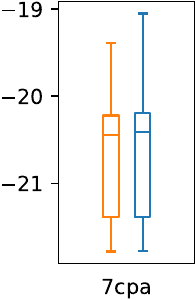} & %
    \includegraphics[height=33mm]{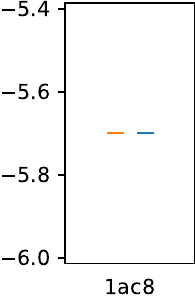} & %
    \includegraphics[height=33mm]{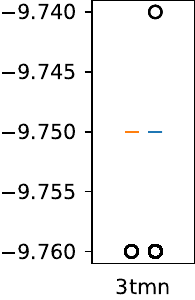} & %
    \includegraphics[height=33mm]{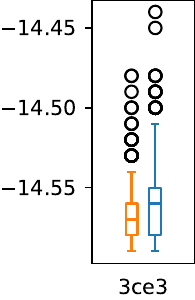}
    \end{tabular*}
    \vspace{-10pt}
    \caption{Distribution of average best energy values for five protein-ligand complexes using the original code, and our method.}
    \vspace{10pt}
    \label{fig:result_whisker_energy}
\end{figure}

Fig.~\ref{fig:result_whisker_energy} presents box-and-whisker plots for the best energy value reached by the scoring function, as reported by AutoDock-GPU. As the initialization process is random, we repeat 1000 runs for each protein-ligand complexes to increase the statistical significance as in~\cite{santos-martins_accelerating_2021}. For each run, the pseudo-random number generator is initialized with the same arbitrary seed for both our code, and the original code.

Table~\ref{tab:error_speedup} reports the absolute and relative errors in the energy value from our method and the AutoDock-GPU baseline. For both \texttt{1ac8} and \texttt{3tmn}, the best energy values show no significant variance between runs for both implementations. For \texttt{1stp}, \texttt{7cpa}, and \texttt{3ce3}, the statistical distribution produced by our code is similar to the one produced by the original code. We notice that for all tested complexes, the relative difference between the average best scores for each method is below $0.18\%$. This observation leads us to conclude that our method provides satisfactory results, and thus validates our approach to perform reduction in the context of AutoDock-GPU.
The justification for this conclusion is two-fold. 
First, the result of the reduction process is used as the energy value, thus a low difference with the reference value shows that our implementation provides a satisfactory level of accuracy for the application. Moreover, the result of the reduction process is used in further computations. Any detrimental error would thus accumulate, and the local-search algorithm would not yield satisfactory results, which is not the case in our tests.

\begin{table}[bt]
    \caption{Absolute difference and relative error in the best energy values and the speedup by our method compared with the baseline.}
    \label{tab:error_speedup}
    \centering
    \begin{tabular}{|c|c|c|c|c|c|}
    \hline
    Complex&\texttt{1stp} &\texttt{7cpa} &\texttt{1ac8} &\texttt{3tmn} &\texttt{3ce3} \\\hline\hline
    $|E_{half} - E_{ref}|$ & $2.00\cdot10^{-5}$ & $3.72\cdot10^{-2}$ & 0.0 & $1.92\cdot10^{-3}$ & $5.78\cdot10^{-3}$\\
    Relative Error & $<0.01\%$ & $0.2\%$ & $0.00\%$ & $0.02\%$ & $0.04\%$\\
    Speedup & $\times 1.16$ & $\times 1.08$ & $\times 1.22$ & $\times 1.27$ & $\times 1.20$\\\hline
    \end{tabular}
\end{table}

\subsection{Runtime Per Evaluation of the Scoring Function}
Next, we evaluate the performance of a single evaluation function. To isolate the reduction process from the energy scoring function, we design a test kernel, where each thread in a block holds a single vector of four single-precision elements. The kernel performs a block-level reduce-and-broadcast operation over all threads. After the reduction operation, the final result is accessible by each thread in their respective local memory. We design two versions of the test kernel.

The first version uses the original AutoDock-GPU code. It first performs a warp-level reduction using warp shuffle functions, which allows to exchange data between threads without using shared memory. %
A block-level reduction is then performed, where the first thread of each warp adds the value it holds to a shared-memory accumulator, using an atomic operation. %
The value of the accumulator is then read back by all threads in the block. %
This three-step process is repeated for each variable that needs to be reduced. The second version of the test kernel uses our TCU-based method.

We measure the elapsed walltime for 1000 launches of each version using the CUDA Runtime API and report the average time. The only parameter influencing the runtime in both versions is the number of threads per block.
64 threads is the lower limit defined by our method -- a 256-element matrix is used to store the values to be reduced, and each threads holds exactly four values, which results in a minimum of 64 threads to fill a single matrix. Future adaptation of the code may overcome this limitation. The upper limit of 1024 is defined by the CUDA platform~\cite{cuda_program}.

\begin{figure}[tb]
    \centering
    \begin{tabular*}{\linewidth}{@{\extracolsep{\fill}} ccc }
    \includegraphics{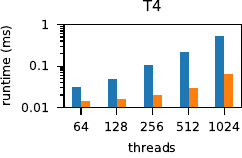} & %
    \includegraphics{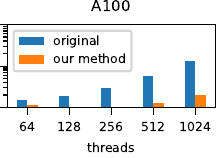} & %
    \includegraphics{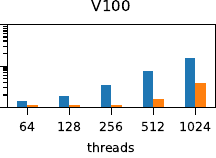}
    \end{tabular*}
    \vspace{-10pt}
    \caption{Average runtime of the two versions of the test reduction kernel on three generations of NVIDIA GPUs: T4, A100, and V100.}
    \vspace{10pt}
    \label{fig:result_runtime}
\end{figure}

Fig.~\ref{fig:result_runtime} shows the average runtime for both versions. The results show that our method consistently performs better than the AutoDock baseline for all block sizes and on all GPUs. This first observation validates the potential of our approach to perform faster block-level reduction in the context of the energy scoring function of AutoDock-GPU. 

We notice that performance for both methods is significantly lower on T4 GPU than on A100 and V100. The lower performance for T4 can be explained by the lower performance Tensor Cores on T4. %
Performance on A100 and V100 are very similar utill the block size of 1024 threads. When using 1024 threads per block, a significant runtime difference is shown on the two GPUs -- the runtime on A100 is 20~ms, which is half of the 39~ms runtime for V100. %
Our profiling results from NVIDIA Nsight Compute show that the test kernel achieved 100\% occupancy on A100 but only 50\% on V100. This low occupancy causes the device to be under-utilized.
Such low theoretical occupancy indicates that the number of active threads per Streaming Multiprocessor is under the maximum achievable value because the resource requirements for the kernel are too high to be accommodated by the device. This could be, for example, the amount of available shared memory.

\begin{figure}
\begin{minipage}[t]{.47\linewidth}
    \includegraphics[width=\columnwidth]{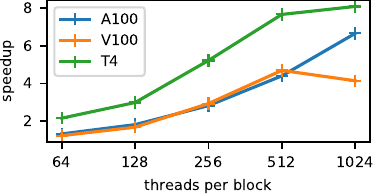}
    \caption{Speedup of the reduction operation using our method over the AutoDock baseline on three GPUs.}
    \label{fig:result_speedup}
\end{minipage}
\hfill
\begin{minipage}[t]{.47\linewidth}
    \includegraphics[width=\columnwidth]{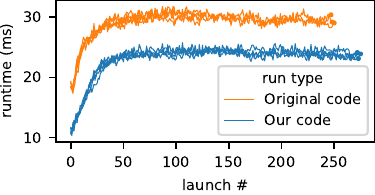}
    \caption{
    Runtime of the local-search kernel, using our TCU-based method and AutoDock-GPU baseline.
    }
    \label{fig:result_profiling_nlaunches}
\end{minipage}
\end{figure}

We evaluate the scalability of our method at increased threads per block. Fig.~\ref{fig:result_speedup} presents the speedup by our reduction method over the baseline on three GPU architectures.
Fig.~\ref{fig:result_profiling_nlaunches} compares the execution times of local search kernel launches during a docking run, using our reduction method or the original method.  %
We observe an increased speedup at an increased number of threads. For instance, the speedup increases from $2\times$ at a block size of 64 on T4 to the maximum of $8.1\times$ on 1024 threads. Overall, the speedup by our method increases linearly with the block size, up to 512 threads per block for all GPUs. 

One interesting observation is that at the maximum block size of 1024 threads, the speedup on A100 increases to a maximum of $6.7\times$ while the speedup on V100 decreases to $4.1\times$. Before reaching the maximum block size, speedup on A100 and V100 GPUs show similar linear scalability.
We investigate this and found from the runtime measurements that the amount of shared memory required when using 1024 threads per block exceeds the hardware limit on V100 GPU, thus resulting in a lower occupancy. Since the original method does not rely on shared memory, this bottleneck only affects our TCU-based method.

\subsection{Impact on the Docking Time}
We evaluate the contribution of our method on the overall simulation. For this, we integrated our block-level reduction method into the scoring function kernel in AutoDock-GPU. We use the docking time, a widely used figure of metric (FoM) in works on AutoDock-GPU~\cite{santos-martins_accelerating_2021,solis-vasquez_benchmarking_2022}. The docking time is reported by AutoDock-GPU, including all docking executions and excluding the I/O operations.%
\begin{figure}[t]
    \centering
    \vspace{10pt}
    \begin{tabular*}{\linewidth}{@{\extracolsep{\fill}} ccccc }
    \multicolumn{5}{c}{\includegraphics[height=4mm]{legend.pdf}} \\
    \includegraphics[height=30mm]{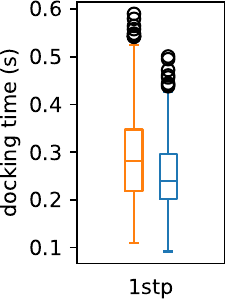} & %
    \includegraphics[height=30mm]{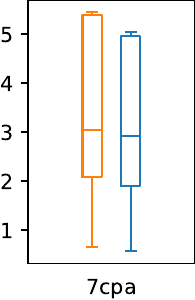} & %
    \includegraphics[height=30mm]{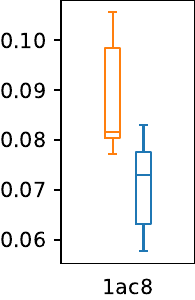} & %
    \includegraphics[height=30mm]{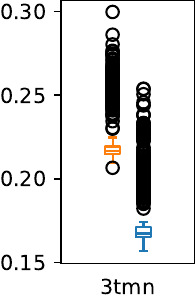} & %
    \includegraphics[height=30mm]{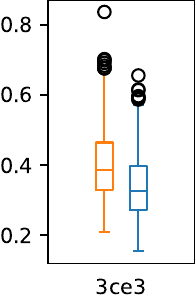}
    \end{tabular*}
    \vspace{-10pt}
    \caption{Docking time on A100 for several protein-ligand complexes, using both the original code and our method.}
    \label{fig:result_whisker_docking_time}
\end{figure}

Fig.~\ref{fig:result_whisker_docking_time} shows the distribution of docking times for five protein-ligand complexes. Note that the docking time is significantly affected by the initial state, which is randomly chosen in AutoDock-GPU. Thus, for a fair comparison, we set the same random initialization seed for both methods. We also gather a large number of samples (1000 runs) to ensure statistical significance of the measurement. %
We observe that our method achieves a lower median, min, max, 25\%, and 75\% percentile docking time compared to the original version. This indicates that our implementation is able to provide consistent speedup over the baseline for general cases.

Distribution of docking times for \texttt{7cpa} exhibits a larger interquartile range compared to the distribution observed for \texttt{1stp}.
This difference is caused by the presence of a significant number of non-convergent runs in the experiments for \texttt{7cpa}.
Non-convergent runs are observed when the search algorithm does not detect convergence, and continues until the maximum number of iterations is reached. This increased iteration count results in significantly higher docking time values for non-convergent runs when compared to convergent ones, for which the search algorithm is stopped earlier.
We measured the proportion of non-convergent runs to be $61\%$ for both versions, when using the \texttt{7cpa} complex.
This indicates that our implementation does not have any impact on convergence of the search algorithm.
Docking runs for other protein-ligand complexes did not exhibit non-convergent runs.

For all test cases, our implementation exhibits a lower average docking time compared to the original code. Table \ref{tab:error_speedup} (row 3) summarizes the speedup by our method over the original AutoDock-GPU code.
We achieved a maximum $\times 1.27$ average speedup, observed for the \texttt{3tmn} complex.
Speedup for the longest-running test case (\texttt{7cpa} complex) is $\times 1.08$.

\section{Related Works}
Molecular docking methods are widely used in drug discovery~\cite{morris_automated_1998,kitchen_docking_2004,stanzione_use_2021}. Various search techniques are used to find the best conformation between molecules \cite{kitchen_docking_2004}, they rely on scoring functions that aim at evaluating the quality of a specific conformation~\cite{stanzione_use_2021}. 
AutoDock is a molecular docking program that relies on a genetic algorithm to find the docking conformation by minimizing a energy-based scoring function~\cite{morris_automated_1998}.

Several works have been conducted to accelerate the original AutoDock code. AutoDock Vina improved AutoDock's local-search method, and made use of multicore and multi-CPU systems to improve performance \cite{trott_autodock_2009}.
AutoDock-GPU added GPU acceleration to AutoDock by adapting the local-search method. Both OpenCL and CUDA versions have been developed. It provided up to a $\times 50$ speedup \cite{santos-martins_accelerating_2021}. 
The recent addition of early stopping to AutoDock-GPU search algorithm allowed to further increase performance~\cite{solis-vasquez_benchmarking_2022}. 
Once adapted for the Summit supercomputer, the CUDA version of AutoDock-GPU allowed to reach a $10\times$ speedup in a real-world docking pipeline \cite{legrand_gpu-accelerated_2020}. Our work proposes a method to increase performance of the CUDA implementation of AutoDock-GPU, by using half-precision number representation in specific portions of the code.

Despite Tensor Cores being specialized in performing operations on small-size matrices, especially for deep learning applications, efforts have been made to make use of this hardware feature to accelerate other applications. For this purpose, algorithms to perform various widely-used operations on Tensor Cores have been developed, such as reduction and scan algorithms~\cite{navarro_gpu_2021,dakkak_accelerating_2019}. In our work, we adapted those methods in order to use them in AutoDock-GPU.
Extensive study of Tensor Cores characteristics have also been conducted. Benchmarking allowed to evaluate Tensor Cores performances in details~\cite{sun_dissecting_2022}. The impact of using half-precision numbers for computation using Tensor Cores, and the associated accuracy loss, have also been documented and precision-refinement techniques have been developed~\cite{markidis_nvidia_2018,haidar2018harnessing}.

\section{Conclusions}
In this work, we investigate a state-of-the-art GPU-accelerated molecular docking software for drug discovery -- AutoDock-GPU. Our profiling results identified a core reduction operation to be sub-optimal due to a large number of synchronization points. We analyzed the specific requirements in the docking process and propose a matrix-based multi-dimensional reduction algorithm for accelerating the local search in AutoDock-GPU. We implemented our method by leveraging NVIDIA Tensor Cores and integrated it in AutoDock-GPU code. We validated our implementation and evaluated its performance on three GPUs. The results show a $4$-$7\times$ speedup of the reduction operation and a 27\% improvement on the average docking time for a real-world docking scenario.

\subsubsection{Acknowledgments} This research is supported by the European Commission under the Horizon project OpenCUBE (GA-101092984).

\bibliographystyle{splncs04}
\bibliography{main}%
\end{document}